\documentclass[prl,twocolumn,showpacs,showkeys,superscriptaddress,citeautoscript]{revtex4-1}
\usepackage{amsmath,amssymb}
\usepackage[colorlinks=true,citecolor=blue,linkcolor=blue]{hyperref}
\usepackage{graphicx}
\usepackage{physics}
\usepackage{tabularx}
\usepackage[capitalise]{cleveref}
\usepackage{xspace}

\newcommand*{\setR}{\ensuremath{\mathbb{R}}}

\newcommand*{\Laplace}{\mathop{}\!\mathbin\bigtriangleup}
\DeclareMathOperator{\del}{\partial}

\newcommand*{\ALGOONE}{\texttt{ALGO1}\xspace}
\newcommand*{\ALGOTWO}{\texttt{ALGO2}\xspace}

\usepackage{acronym}
\newacro{cft}[\textsc{cft}]{Conformal field theory}
\newacroplural{cft}[\textsc{cft}s]{Conformal field theories}

\newacro{qft}[\textsc{qft}]{quantum field theory}
\newacro{eft}[\textsc{eft}]{effective field theory}
\newacroplural{eft}[\textsc{eft}s]{effective field theories}

\newacro{vev}[\textsc{vev}]{vacuum expectation value}
\newacro{rg}[\textsc{rg}]{renormalization group}
\newacro{dof}[\textsc{dof}]{degree of freedom}
\acrodefplural{dof}[\textsc{dof}]{degrees of freedom}
\newacro{ir}[\textsc{ir}]{infrared}
\newacro{eom}[\textsc{eom}]{equation of motion}
\acrodefplural{eom}[\textsc{eom}]{equations of motion}

\newacro{qcd}[\textsc{qcd}]{quantum chromodynamics}

\newacro{mc}[\textsc{mc}]{Monte Carlo}

\newcommand*{\Left}{\textsc{l}}
\newcommand*{\Right}{\textsc{r}}
\newcommand*{\setZ}{\mathbb{Z}}

\usepackage{microtype}

\begin{document}

\title{Conformal dimensions in the large charge sectors at the $O(4)$
  Wilson--Fisher fixed point}

\author{Debasish Banerjee}
\email{debasish.banerjee@phyik.hu-berlin.de}
\affiliation{Institut f\"ur Physik, Humboldt-Universit\"at zu Berlin, Zum Gro\ss en Windkanal 6, D-12489 Berlin, Germany}
\author{Shailesh Chandrasekharan}
\email{sch@phy.duke.edu}
\affiliation{Department of Physics, Duke University, Durham, North Carolina 27708, \textsc{usa}}
\author{Domenico Orlando}
\email{domenico.orlando@to.infn.it}
\affiliation{\textsc{infn}, sezione di Torino and Arnold–Regge Center, via Pietro Giuria 1, 10125 Torino, Italy}
\affiliation{Albert Einstein Center for Fundamental Physics, Institute for Theoretical Physics, University of Bern,
Sidlerstrasse 5, \textsc{ch}-3012, Bern, Switzerland}
\author{Susanne Reffert}
\email{sreffert@itp.unibe.ch}
\affiliation{Albert Einstein Center for Fundamental Physics, Institute for Theoretical Physics, University of Bern, Sidlerstrasse 5, \textsc{ch}-3012, Bern, Switzerland}

\begin{abstract}
  We study the $O(4)$ Wilson--Fisher fixed point in $2+1$ dimensions in fixed large-charge sectors identified by products of two spin-$j$ representations $(j_{\Left},j_{\Right})$.
  Using effective field theory we derive a formula for the conformal dimensions $D(j_{\Left},j_{\Right})$ of the leading operator  in terms of two constants, $c_{3/2}$ and $c_{1/2}$, when the sum $j_{\Left} + j_{\Right}$ is much larger than the difference $\abs{j_{\Left}-j_{\Right}}$. We compute $D(j_{\Left},j_{\Right})$ when $j_{\Left}= j_{\Right}$ with \acl{mc} calculations in a discrete formulation of the $O(4)$ lattice field theory, and show excellent agreement with the predicted formula and estimate $c_{3/2}=1.068(4)$ and $c_{1/2}=0.083(3)$.
\end{abstract}
\preprint{}
\maketitle

\section{Introduction}%
\label{sec:intro}

\ac{cft} holds a central place in the study of \ac{qft}, as it is relevant to both particle physics and condensed matter systems at criticality, and \emph{via} the gauge/gravity correspondence even to the description of quantum gravity.
Generically, \acp{cft} do not contain any small couplings that can be used in a perturbative analysis.
However, the conformal symmetry constrains its observables such that we can determine any \(n\)-point function using only operator dimensions and three-point function coefficients.
While it is possible to treat strongly coupled theories with methods such as the large-$N$ expansion, the $\epsilon$-expansion (see~\cite{Pelissetto:2000ek} for a review) and the conformal bootstrap~\cite{Rattazzi:2008pe}, they are notoriously difficult to access analytically. In simple cases, \ac{mc} techniques offer a reliable numerical alternative~\cite{Campostrini:2000iw,Campostrini:2002ky}.

Recently, it has been shown in a series of papers~\cite{Hellerman:2015nra,Alvarez-Gaume:2016vff,Monin:2016jmo,Loukas:2017lof,Loukas:2018zjh,Favrod:2018xov} that working in a sector of large global charge results in important simplifications and gives us a perturbative handle to study \acp{cft} using \acp{eft}: it is possible to write an effective action as an expansion in terms of a large conserved charge with unknown coefficients.
For the Wilson--Fisher point in the three-dimensional $O(N)$ vector model~\cite{Wilson:1971dc}, except for two low-energy couplings, all terms are suppressed by inverse powers of the large charge~\cite{Alvarez-Gaume:2016vff}.
The approximate physics of the \ac{cft} becomes accessible as a function of these two couplings which we label as $c_{3/2}$ and $c_{1/2}$.
This suggests a double-pronged approach to \acp{cft}, which involves using the large-charge expansion to determine the effective action, paired with \ac{mc} calculations to determine the low-energy couplings.
For the case of the \(O(2)\) Wilson--Fisher \ac{cft}, this approach has been successfully implemented recently~\cite{PhysRevLett.120.061603}. In particular, it was shown that the predictions obtained with the two couplings remain very accurate even for low charges.

In this letter, we explore the viability of this approach for the \(O(4)\) Wilson--Fisher \ac{cft}, which has qualitatively distinct features from the $O(2)$ model studied earlier. The fact that $O(4)$ symmetry is non-Abelian and that it leads to two conserved global charges $j_\Left$ and $j_\Right$, creates novel challenges. The ground state can become spatially inhomogeneous requiring a different analysis in the \ac{eft}, and the construction of a worldline-based lattice model becomes necessary to access easily the large-charge sectors.
The \ac{cft} with $O(4)$ symmetry is also interesting in many subfields of physics. For example, it arises naturally in the study of finite-temperature chiral phase transitions in two-flavor \ac{qcd} with massless quarks~\cite{PhysRevD.29.338,Wilczek:1992sf}. It is also of interest in studies of strongly correlated electronic systems at half filling built out of models of interacting electrons with spin~\cite{Yang90}.

\begin{figure}
\includegraphics [width=\linewidth]{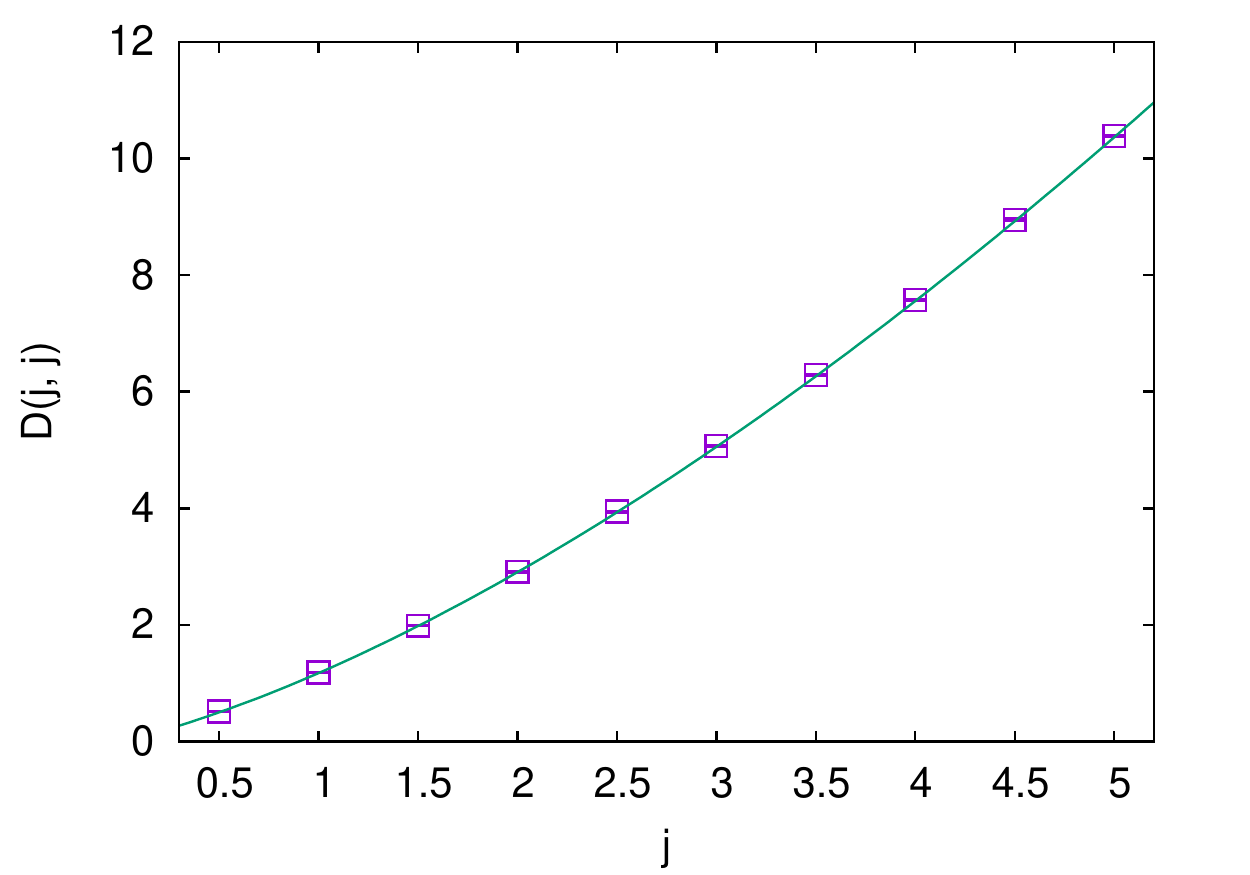}
\caption{Plot of $D(j,j)$ as a function of $j$. The squares represent the data obtained using \ac{mc} calculations with the lattice model in~\cref{eq:fo4model}.
  The solid line is the large-charge prediction~\cref{eq:dimensions-jj} with $c_{3/2}=1.068(4)$ and $c_{1/2}=0.083(3)$. \label{fig:anomD}}
\end{figure}
Traditional $O(4)$ lattice models are constructed using classical vectors. Unfortunately, in the study of large charge sectors, using traditional \ac{mc} methods based on sampling classical vectors leads to severe signal-to-noise ratio problems. While worldline representations can in principle solve these problems~\cite{Wolff:2010zu,Gattringer:2017hhn}, the presence of an infinite Hilbert space at each lattice site can still lead to algorithmic inefficiencies. Fortunately, a discrete version of the $O(4)$ model with a finite Hilbert space per lattice site is easy to construct~\cite{PhysRevD.77.014506,Chandrasekharan:2010ik}.
In this work we use this discrete formulation to accurately compute the conformal dimensions $D(j_\Left,j_\Right)$ (defined below) at the $O(4)$ Wilson--Fisher fixed point, when $j_\Left=j_\Right=j$ (see Figure~\ref{fig:anomD}). The large-charge prediction (see~\cref{eq:dimensions-jj}) is an excellent fit to the lattice data even up to the smallest charge giving us $c_{3/2}=1.068(4)$ and $c_{1/2}=0.083(3)$.

\section{Large charge predictions}\label{sec:large-charge}

The \ac{eft} approach to the $O(4)$ \ac{cft} is based on the construction of an effective action at large charge.
Using the fact that $SU_{\Left}(2) \times SU_{\Right}(2)$ is a double cover of $O(4)$, this action can be put into the form~\cite{Alvarez-Gaume:2016vff,Loukas:2017lof}
\begin{equation}
  \label{eq:large-charge-action}
  S = \int_{\setR \times \Sigma} \dd{t} \dd{\Sigma} \Big[ \frac{\sqrt{2}}{27 c_{3/2}^2} \norm{\dd g}^3  - \frac{ c_{1/2}}{3 \sqrt{2}c_{3/2}} R \norm{\dd g} + \dots  \Big],
\end{equation}
where \(g(\mathbf{r},t) \in SU(2)\), \(\mathbf{r} \) is the coordinate on \(\Sigma\), \(\norm{\dd g}^2 = \Tr( \del_\mu g^\dagger \del^\mu g)\), and \(c_{3/2}\) and \(c_{1/2}\) are the two leading low-energy couplings referred to earlier.
This action is to be understood as an expansion around the fixed-charge ground state. We study the system on a spatial Riemann surface \( \Sigma\) with scalar curvature \(R\). %
The field \(g\) transforms as \(g \to V_{\Left} g V_{\Right}^{-1}\) under the global \(SU(2)_{\Left} \times SU(2)_{\Right}\). The corresponding Noether charges are the two-by-two Hermitian traceless matrices in the \(su(2)\) algebra:
\begin{align}
  Q_{\Left} &= i \int \dd{\Sigma } c_J \del_0 g g^\dagger, & Q_{\Right} &= i \int \dd{\Sigma } c_J  \del_0 g^\dagger g,
\label{eq:NoetherQ}
\end{align}
where \(c_J =\sqrt{2}  \norm{\dd g} /(9 c_{3/2}^2) - c_{1/2} R/ \pqty{6 \sqrt{2} c_{3/2}\norm{\dd g}} \). Their two eigenvalues are $\pm j_{\Left} $ and $\pm j_{\Right}$. Under the action of \(SU(2)_{\Left} \times SU(2)_{\Right}\) the charges in \cref{eq:NoetherQ} transform as
\begin{align}
  Q_{\Left} &= V_{\Left} Q_{\Left} V_{\Left}^{-1 }, &   Q_{\Right} &= V_{\Right} Q_{\Right} V_{\Right}^{-1 },
\end{align}
but $j_{\Left} $ and $j_{\Right}$ remain invariant. We will refer to the class of configurations $g$ connected by \(SU(2)_{\Left} \times SU(2)_{\Right}\) transformations as the $(j_\Left,j_\Right)$ sector. In the underlying \ac{qft} the sectors $(j_\Left,j_{\Right})$ naturally label the irreducible representation space of \(SU(2)_{\Left} \times SU(2)_{\Right}\), hence the values of $j_\Left$ and $j_\Right$ are quantized, \emph{i.e.}, \(j_\Left,j_{\Right} \in \tfrac{1}{2} \setZ \). Their sum must be integer, \(j_{\Left} + j_{\Right} \in \setZ \), because we consider only states that are representations of \(O(4)\).

Instead of $Q_{L,R}$ it is more convenient to work with the projections \(q_{\Left, {\Right}} = \Tr(Q_{\Left, {\Right}} \sigma_3)/2\).
In a fixed $(j_\Left,j_\Right)$ sector these projections will take values in the range \(-j_{\Left, {\Right}}\leq q_{\Left,\Right}\leq  j_{\Left, {\Right}}\).
It is natural to identify them with the quantized charges of the states in the representation with highest weights \((j_\Left,j_\Right)\).
In Ref.~\cite{Alvarez-Gaume:2016vff} it was shown that the minimal energy solutions to the \ac{eom} for the action~\eqref{eq:large-charge-action} for fixed values of \(q_{\Left, \Right}\) are homogeneous in space and arise in sectors with \(j_\Left = j_\Right = \max(q_\Left, q_\Right)\).
This leads to the formula for the minimal energy in a fixed \((j,j)\) sector:
\begin{equation}
  \label{eq:energy-j-j}
  E(j, j) = \sqrt{\frac{8j^3}{V}} \pqty{ c_{3/2} +  c_{1/2} \frac{R V}{4j} + \dots } + \zeta(-1/2| \Sigma),
\end{equation}
where \(\zeta(s| \Sigma)\) is the \(\zeta \)--function for the Laplacian on the surface \(\Sigma\) and represents the contribution of the Casimir energy.
Since in a \ac{cft} the conformal dimension of an operator is identified with the energy on the unit sphere \(\Sigma = S^2\) of the corresponding state, we deduce a formula for the dimension of the lowest operator in the $(j,j)$ sector:
\begin{equation}
  \label{eq:dimensions-jj}
  D(j,j) = \sqrt{\frac{2j^3}{\pi}} \pqty{ c_{3/2} + c_{1/2} \frac{2\pi}{j} + \order{\frac{1}{j^2}} } + c_0 ,
\end{equation}
where $c_0 = \zeta(-1/2| S^2) \approx -0.094 $ is a universal constant~\cite{Elizalde:2012zza,delaFuente:2018qwv}.

In this work we generalize this formula to any representation \((j_{\Left}, j_{\Right})\).
We need to find the minimal-energy solutions admitted by the action Eq.~\eqref{eq:large-charge-action} whose charge matrices \(Q_{\Left}\) and \(Q_{\Right}\) are diagonal and correspond, by the argument above, to highest-weight states in a representation of \(SO(4)\).  In order to study the sectors with $j_\Left \neq j_\Right$, the analysis in Refs.~\cite{Alvarez-Gaume:2016vff,Hellerman:2017efx,Hellerman:2018sjf} suggests that we need to look for inhomogeneous field configurations
\(g(\mathbf{r},t)\) of the form
\begin{equation}
  g(\mathbf{r},t) =
  \begin{pmatrix}
    \cos(p(\mathbf{r})) e^{i \mu_1 t} & \sin(p(\mathbf{r})) e^{i \mu_2 t} \\
     -\sin(p(\mathbf{r})) e^{-i \mu_2 t} & \cos( p(\mathbf{r})) e^{-i \mu_1 t}\\
  \end{pmatrix},
\end{equation}
where \(\mu_1\) and \(\mu_2\) are constants parameterizing the action of \(SU(2)_{\Left}\times SU(2)_{\Right}\) on the inhomogeneous configuration encoded by the undetermined function $p(\mathbf{r})$. The homogeneous solution of Ref.~\cite{Alvarez-Gaume:2016vff} with \(\mu_1 = \mu_2 = \order{j^{1/2}} \) and \(p(\mathbf{r}) = 0\) describes the $j_\Left=j_\Right$ sectors.
In order to explore solutions with small non-zero values of $\abs{j_\Left - j_\Right}/\max(j_{\Left}, j_{\Right})$ we may expand the action in a series in \(\eta^2 = \pqty{\mu_2 - \mu_1}/\pqty{\mu_2 + \mu_1} \) which will naturally be small.
At leading order in \(\eta^2\)
the \ac{eom} is an elliptic sine--Gordon equation
\begin{equation}
  \label{eq:sine-Gordon}
  2 \Laplace p(\mathbf{r}) + \Lambda^2 \sin( 2 p (\mathbf{r})) = 0, 
\end{equation}
where \(\Laplace\) is the Laplacian on \(\Sigma\) and \(\Lambda^2 = \mu_2^2 - \mu_1^2\).
The case \(\Sigma = T^2\) was already discussed in~\cite{Hellerman:2017efx,Hellerman:2018sjf}.
Here we concentrate on \(\Sigma = S^2(r_0)\) in order to calculate operator dimensions.
We express the parameters \(\mu_1\) and \(\mu_2\) as functions of the eigenvalues \(j_{\Left}\) and \(j_{\Right}\) and find that the leading (tree-level) contribution to the energy of the solutions to the \ac{eom} has the same form as Eq.~\eqref{eq:energy-j-j}, where now \(j = j_m = \max(j_{\Left}, j_{\Right})\), plus an extra contribution that captures the inhomogeneity of the solution  (\emph{i.e.}, $\nabla p(\mathbf{r}) \neq 0$):
\begin{equation}
  E_{\Sigma}^{\text{tr}} = \sqrt{\frac{8j_m^3}{V}} \bqty{ c_{3/2} + \frac{c_{1/2}  RV}{4j_m}  +  \int_\Sigma \dd{\Sigma} \frac{(\nabla p)^2}{6 c_{3/2} j_m}  + \dots }.
\end{equation}
As discussed in the supplementary material, when \(\Sigma = S^2(r_0)\), the \ac{eom}~\eqref{eq:sine-Gordon} admits different branches of smooth solutions, parameterized by an integer \(\ell\) which counts the zeros of \(p(\mathbf{r})\).
The energy is minimal in the first non-trivial branch (\(\ell = 1\)), where \(2 \le r_0^2 \Lambda^2 < 6\).
Here the integral of the divergence can be computed numerically in terms of an expansion in \(\abs{j_\Left - j_\Right}/ j_m \) to give
\begin{equation}
  \label{eq:kinetic-term}
  \frac{1}{4 \pi} \int_{S^2} \dd{\Omega} (\nabla p)^2 =  \frac{\abs{j_{\Left} - j_{\Right}}}{j_m } + \lambda_2 \pqty{\frac{\abs{j_{\Left} - j_{\Right}}}{j_m }}^2 + \dots %
\end{equation}
with \(\lambda_2 \approx 0.2455 \). %
This is the leading contribution in the large-charge expansion.
There will be in general higher-order corrections suppressed by inverse powers of the large charges due to sub-leading terms in the tree-level action in Eq.~\eqref{eq:large-charge-action} and to quantum corrections.

There is only one term of order \(\order{j^0}\): the Casimir energy of the Goldstones resulting from the spontaneous symmetry breaking \(SO(3) \times D \times SO(2)^2 \to SO(2) \times D'\) discussed in the supplementary material.
The two broken generators of the isometries on the sphere only give rise to one Goldstone \ac{dof}. Together with the 2 \ac{dof} from the broken internal symmetries, they are arranged into one type-I and one type-II Goldstone field in the notation of~\cite{Nielsen:1975hm}.
Only the former contributes to the Casimir energy as \(E_0 = \zeta(-1/2| S^1)/(2\sqrt{2}) \). The zero-point energy is different from the one in the \((j,j)\) sector because the low-energy excitations only propagate in the direction of the unbroken sphere isometry.
Once again we can use the state/operator correspondence and obtain the final formula for the conformal dimension of the lowest operator in the representation \((j_{\Left}, j_{\Right})\) of \(SO(4)\) when $ j_{\Left} \neq j_{\Right}$:
\begin{widetext}
\begin{align}
  \label{eq:dimensions-general-irrep}
  D(j_{\Left},j_{\Right}) &= \sqrt{\frac{2j_m^3}{\pi}} \bqty{ c_{3/2} +  c_{1/2} \frac{2 \pi}{j_m} + \frac{1}{3 c_{3/2}} \pqty {\frac{\abs{j_{\Left} - j_{\Right}}}{j_m} +  \lambda_2 \frac{\pqty{j_{\Left} - j_{\Right}}^2}{j_m^2 }  + \dots } \frac{2 \pi}{ j_m } + \dots } - \frac{1}{12 \sqrt{2}}.  
\end{align}
\end{widetext}
As we have stressed, the conformal dimensions only depend on the two Wilsonian couplings \(c_{3/2}\) and \(c_{1/2}\), which are the same coefficients that appear in Eq.~\eqref{eq:dimensions-jj} for the $j_\Left=j_\Right$ case. We now explain how we  determine them using \ac{mc} methods with our lattice model.

\begin{figure}[h]
\includegraphics [width=0.8\linewidth]{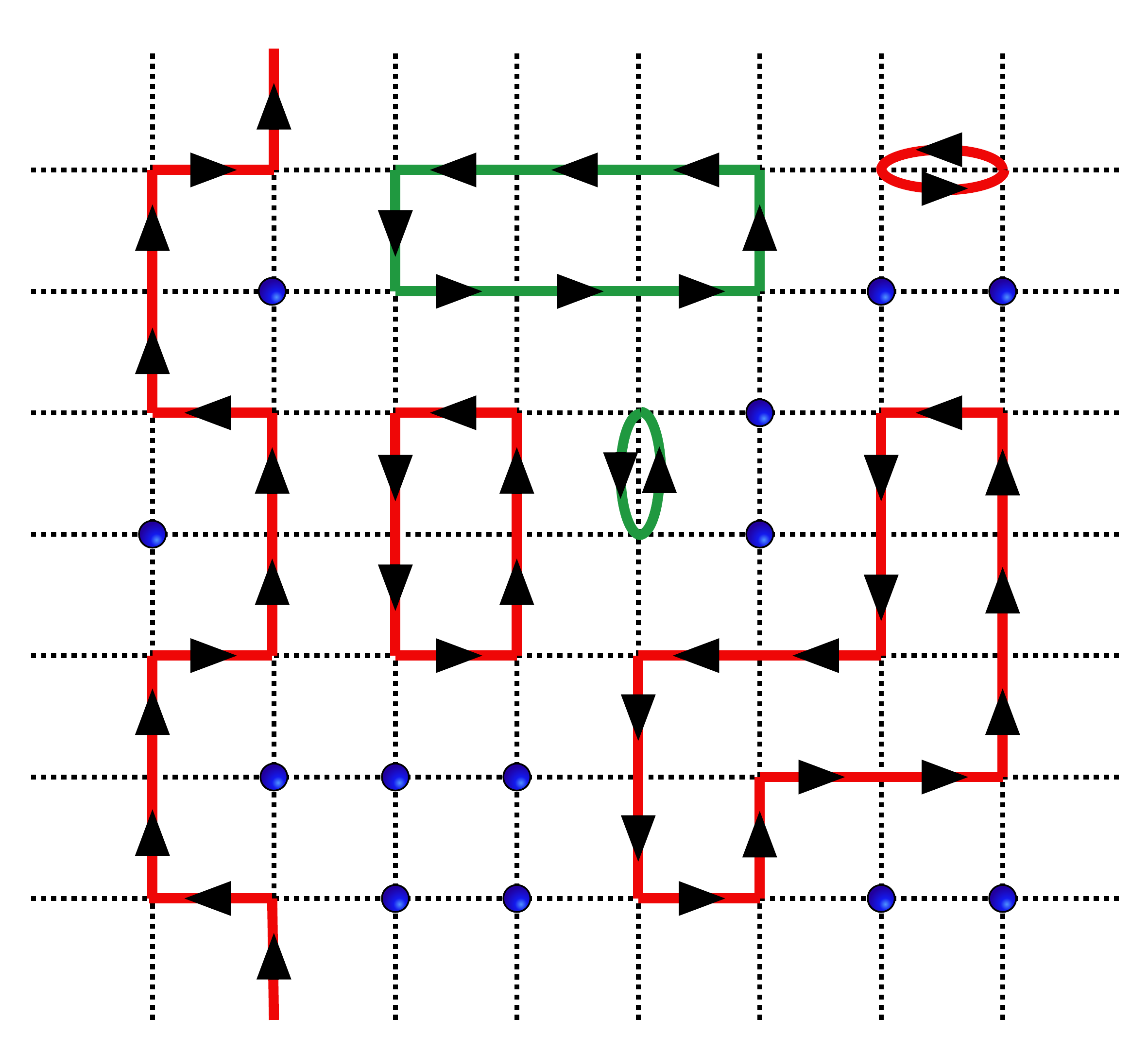}
\caption{Illustration of an $O(4)$ worldline configuration in two dimensions. The solid circles represent vacuum sites, each of which have a weight $U$. All other sites have a single $O(4)$ particle with charge $(q_\Left,q_\Right)=(\pm 1/2,\pm 1/2)$ moving in space-time.\label{fig:wlconf}}
\end{figure}

\section{Lattice simulations}\label{sec:lattice}

Our lattice model was first introduced in Ref.~\cite{PhysRevD.77.014506} as a model for pion physics in two-flavor \ac{qcd} and studied with an efficient \ac{mc} algorithm. It is constructed using four Grassmann fields $\psi_\alpha(x), \overline{\psi}_\alpha(x), \alpha=1,2$ at every three-dimensional periodic cubic lattice site $x = (\mathbf{r},t)$ of size $L$ in all the directions. If we arrange these four-fields into a $2\times 2$ matrix of the form $g_{\alpha \beta}(x) = \psi_\alpha \overline{\psi}_\beta$ we can write the lattice action as
\begin{align}
  S  = - \sum_{\ev{xy}} \Tr(g_x g_y) - \frac{U}{2} \sum_x \det(g_x),
\label{eq:fo4model}
\end{align}
where $\ev{xy}$ are nearest-neighbor bonds. This action is invariant under the $SU(2)\times SU(2)$ transformations $g_x \rightarrow V_\Left g_x V^{-1}_\Right$ on odd sites and $g_x \rightarrow V_\Right g_x V^{-1}_\Left$ on even sites. The partition function of the model can be expressed as a sum over configurations where each site either contains a vacuum site or a worldline of an $O(4)$ particle in the vector representation. Thus, each worldline has four possible states that label the eigenvalues $(q_\Left,q_\Right) = (\pm 1/2,\pm 1/2)$ of particles that travel through the sites. These can be thought of as oriented loops with two colors (say red and green). An illustration of a configuration is shown in~\cref{fig:wlconf}.

\begin{figure}
\includegraphics [width=0.8\linewidth]{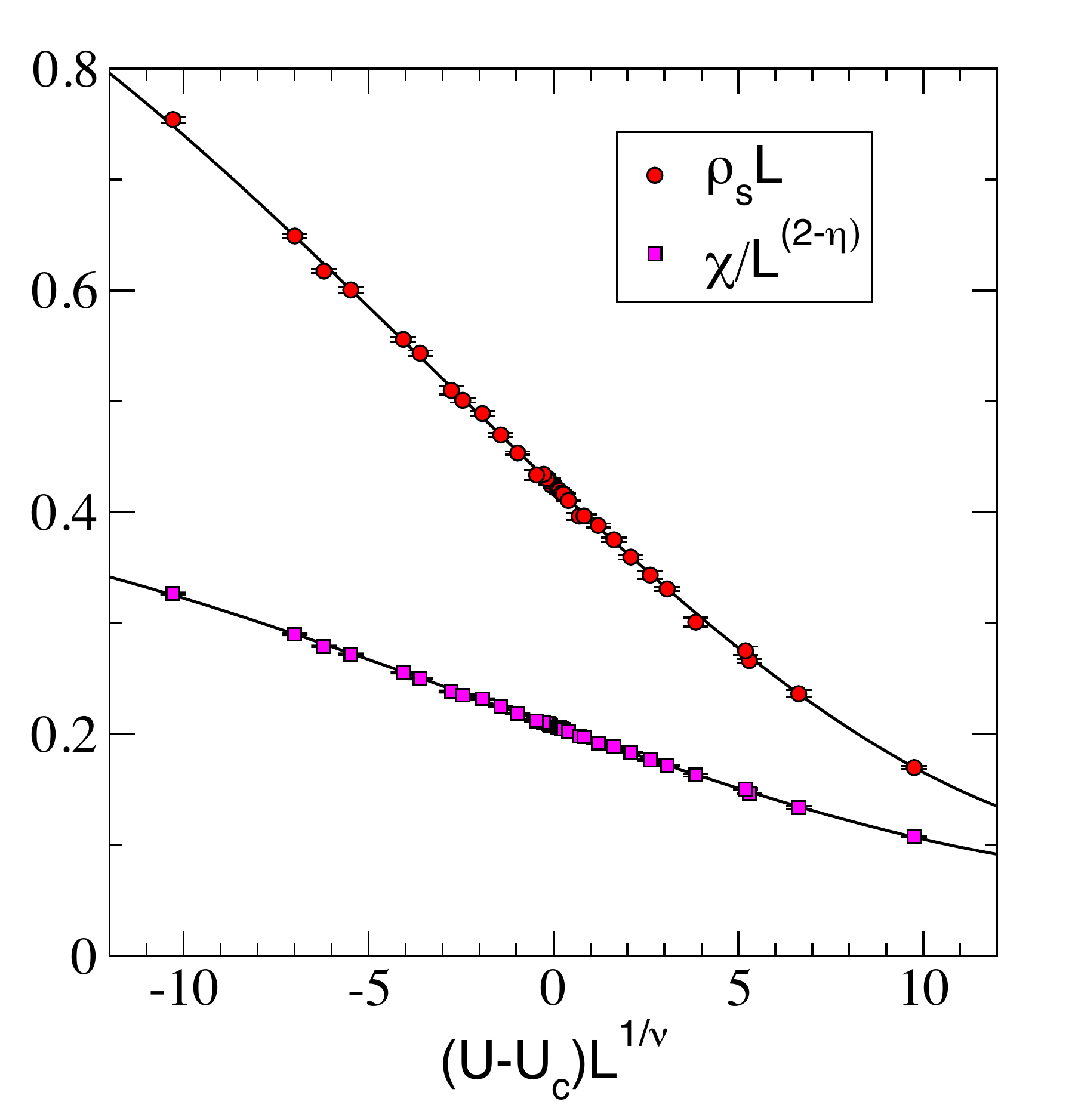}
\caption{The critical scaling plots of $\rho_s L$ (circles) and $\chi L^{2-\eta}$ as a function of the scaling variable $(U-U_c)L^{1/\nu}$. The solid lines show the goodness of the combined fit of all the data shown to polynomials to fourth order.\label{fig:critdata}}
\end{figure}

The weight of a worldline configuration is given by $U^{N_m}$ where $N_m$ is the number of vacuum sites. As $U$ is tuned, the model undergoes a phase transition between the massive symmetric phase at large values to a phase where the $O(4)$ symmetry is spontaneously broken at small values. Using well-established \ac{mc} methods~\cite{PhysRevD.77.014506, Chandrasekharan:2010ik} we first demonstrate that at the critical point we obtain the $O(4)$ Wilson--Fisher \ac{cft} by computing the critical exponents $\nu$ and $\eta$. For this purpose we compute the current susceptibility $\rho_s$ and the order parameter susceptibility $\chi$, details of which can be found in the Supplementary Material. Finite-size scaling theory near a second-order phase transition predicts that $\rho_s L$ and $\chi L^{2-\eta}$ must be simple polynomials of $(U-U_c)L^{1/\nu}$. A combined fit of our data gives $U_c=1.655394(3)$, $\nu=0.746(3)$ and $\eta=0.0353(10)$. In~\cref{fig:critdata} we plot our data and the fit. These exponents are in excellent agreement with earlier results, $\nu=0.749(2)$ and $\eta=0.0365(10)$, obtained from the traditional lattice model~\cite{Hasenbusch:2000ph}.

Having established that our lattice model indeed reproduces the $O(4)$ \ac{cft} when $U=U_c$, we can use the method we developed in Ref.~\cite{PhysRevLett.120.061603} to accurately compute the conformal dimensions $D(j,j)$ at the $O(4)$ \ac{cft}. We can create configurations in a specific $(j_\Left,j_\Right)$ sector by placing appropriately charged sources and sinks at $t=0$ and $t=L/2$ respectively. More concretely, sources that create a red loop are assigned the charge $(1/2,1/2)$ and the sinks that annihilate them are assigned the charge $(-1/2,-1/2)$. Similarly, those that create and annihilate the green loops are assigned charges $(1/2,-1/2)$ and $(-1/2,1/2)$. Using these fundamental sources we can construct sources and sinks with any charge $(q_\Left,q_\Right)$. However, since each site can only have one red or one green source, to create a source with a large charge we distribute the fundamental sources in a local region near the origin (see Supplementary Material for more details). 
Since the couplings $c_{3/2}$ and $c_{1/2}$ can be computed by fitting the data for $D(j,j)$ to the predicted form in \eqref{eq:dimensions-jj}, in this work we only study the sector with $j_\Left=j_\Right=j, j=1/2,1,3/2,\dots$. 
For this purpose we only work with sources and sinks of equal charges by creating $2j$ sources of red loops at $t=0$ and annihilating them at $t=L/2$. This naturally projects us into the highest-weight representation sector with $j_\Left=j_\Right=j$. Let $Z_j(L)$ be the partition function in the presence of these sources and sinks. In Ref.~\cite{PhysRevLett.120.061603} we developed an efficient algorithm to compute the ratio $R_j(L) = Z_{j}(L)/Z_{j-1/2}(L)$, which is expected to scale as  $C/L^{2\Delta(j)}$ for large values of $L$. By evaluating $R_j(L)$ for various values of $j,L$ and fitting to the expected form we can accurately compute the difference in the conformal dimensions $\Delta(j) = D(j,j)-D(j-1/2,j-1/2)$. From these differences we can also estimate $D(j,j)$, since conformal invariance fixes $D(0,0)=0$. Our final results are tabulated in~\cref{tab:conformal_dimensions} up to $j=5$. As the table shows, our results are also in good agreement with earlier calculations up to $j=2$~\cite{PhysRevB.84.125136}. 
\begin{table}
\begin{tabular}{|c|c|c||c|c|c|}
\hline
$j$ & \multicolumn{2}{c||}{$D(j,j)$} & 
$j$ & \multicolumn{2}{c|}{$D(j,j)$}                                                                                          \\
    & (this work)                    & (from~\cite{PhysRevB.84.125136}) &   & (this work) & (from~\cite{PhysRevB.84.125136}) \\ 
\hline 
1/2 & 0.515(3)                       & 0.5180(3)                        & 1 & 1.185(4)    & 1.1855(5)                        \\
3/2 & 1.989(5)                       & 1.9768(10)                       & 2 & 2.915(6)    & 2.875(5)                         \\
5/2 & 3.945(6)                       & -                                & 3 & 5.069(7)    & -                                \\
7/2 & 6.284(8)                       & -                                & 4 & 7.575(9)    & -                                \\
9/2 & 8.949(10)                      & -                                & 5 & 10.386(11)  & -                                \\
\hline
\end{tabular}
\caption{\label{tab:conformal_dimensions} Results for the conformal dimensions $D(j,j)$ up to $j=5$ computed using worldline \ac{mc} methods in this work (second and fifth column). We also compare our results with earlier calculations up to $j=2$ found in~\cite{PhysRevB.84.125136}.}
\end{table}
Fitting the data in~\cref{tab:conformal_dimensions} to the large $j$ form in~\cref{eq:dimensions-jj} we obtain $c_{3/2} = 1.068(4)$ and $c_{1/2} = 0.083(3)$ (see~\cref{fig:anomD}).

\section{Conclusions}\label{sec:conclusions}

In this letter we provide a new prediction for the anomalous dimensions $D(j_{\Left},j_{\Right})$ (see~\eqref{eq:dimensions-general-irrep}) at the $O(4)$ Wilson--Fisher fixed point in terms of the two couplings that appear in the fixed large-charge effective action~\eqref{eq:large-charge-action}. Our prediction is valid in the limit of large $(j_{\Left},j_{\Right})$ and small $\abs{j_\Left-j_\Right}/\max(j_{\Left}, j_{\Right})$. We then use a discrete lattice $O(4)$ model to compute the two couplings by fitting the data for $D(j,j)$ to the prediction in~\cref{eq:dimensions-jj} obtained from an earlier work. We also demonstrate that this prediction provides an excellent approximation even at small values of $j$ (see~\cref{fig:anomD}). Our  estimate $c_{3/2}=1.068(4)$ and $c_{1/2}=0.083(3)$ can be used in~\eqref{eq:dimensions-general-irrep} to predict $D(j_{\Left},j_{\Right})$ even for $j_\Left \neq j_\Right$. While our lattice model can in principle be used to check the validity of these predictions, our method is likely to suffer from signal to noise ratio problems when $j_\Left$ and $j_\Right$ are sufficiently large and different.
Discrete lattice models like ours can in principle also be designed for other non-Abelian symmetry groups, thus allowing us to explore the robustness of the large-charge \ac{eft} method for general \acp{cft}. Such extensions are likely to bring new challenges providing a fertile ground for further research.

\section*{Acknowledgments}

We wish to thank Andrew Gasbarro, Simeon Hellerman, Francesco Vitali, Masataka Watanabe, Urs Wenger and Uwe--Jens Wiese for valuable discussions.
The material presented here is based upon work supported by the U.S. Department of Energy, Office of Science, Nuclear Physics program under Award Number DE-FG02-05ER41368. D.B. is supported by the German Research Foundation (DFG), Grant ID BA 5847/2-1.
D.O. acknowledges partial support by the NCCR 51NF40-141869 ``The Mathematics of Physics'' (SwissMAP). 
The work of S.R. is supported by the Swiss National Science Foundation (\textsc{snf}) under grant number PP00P2\_157571/1.

\bibliography{LargeQ}

\clearpage

\section{Supplementary Material}
\subsection{Linearized EOM and symmetry breaking pattern}

To our knowledge, there is no known solution to the \ac{eom} in Eq.~\eqref{eq:sine-Gordon} in terms of elementary functions.
Nonetheless, we can still understand some of its qualitative properties via the study of the associated linear problem, obtained by setting \(p(\mathbf{r}) = \epsilon^2 \varpi(\mathbf{r})\) and taking the limit \(\epsilon \to 0\).
At leading order, \(\varpi\) satisfies the Laplace equation:
\begin{equation}
  \Laplace \varpi(\mathbf{r}) + \Lambda^2 \varpi(\mathbf{r}) = 0  .
\end{equation}
As it is well known, when \(\Sigma = S^2(r_0)\) this equation admits smooth solutions only for \(\Lambda^2 r_0^2 = \ell(\ell + 1)\), \(\ell = 0, 1, \dots\).
These are the spherical harmonics \(\varpi(\theta, \phi) = Y_{\ell m}(\theta, \phi)\).
Imposing reality for \(\varpi\) selects \(m = 0\).
Moreover, we know that energy minimization requires \(\Lambda^2\) to take the smallest non-vanishing value \(\Lambda^2 = 2/r_0^2\).
The final solution in a convenient normalization is then
\begin{equation}
  \varpi(\theta) = Y_{1,0}(\theta) = \sqrt{\frac{3}{4 \pi}} \cos(\theta) .
\end{equation}
In the non-linear equation~\eqref{eq:sine-Gordon}, the values of \(\Lambda\) are not quantized, but for each value of \(\ell\) there is a continuous branch of solutions with \(\ell (\ell + 1) \le \Lambda^2 r_0^2 < (\ell + 1) (\ell + 2) \).

It is convenient to express \(\epsilon^2\) as function of the charges.
At leading order in \(\eta\),
\begin{equation}
  \frac{\abs{j_{\Left} - j_{\Right}}}{2 j_m} = \frac{1}{V} \int_{S^2} \dd{\Omega} \sin[2](p) = \frac{\epsilon^2}{4 \pi r_0^2} \int_{S^2} \dd{\Omega} \varpi(\mathbf{r})^2 = \frac{\epsilon^2}{4 \pi},
\end{equation}
which shows that the linear limit \(\epsilon^2 \to 0\) is the same as requiring \(\abs{j_{\Left} - j_{\Right}} \ll \max(j_{\Left}, j_{\Right})\).
The leading contribution to the integral appearing in the energy formula~\eqref{eq:kinetic-term} is then:
\begin{multline}
  \int_{S^2} \dd{\Omega} (\nabla p)^2 = - \int_{S^2} \dd{\Omega} p \Laplace p = \Lambda^2 \int_{S^2} \dd{\Omega} p^2(\mathbf{r}) \\
  = \Lambda^2 r_0^2 \epsilon^2 = 2 \epsilon^2 = 4 \pi \frac{\abs{j_{\Left} - j_{\Right}}}{ j_m}.
\end{multline}

This explicit solution lets us identify the symmetry-breaking pattern associated to fixing the two representations independently.
The minimal-energy configuration is not homogeneous, but a function of the polar angle \(\theta\).
This breaks the \(SO(3)\) isometry of the sphere spontaneously to \(SO(2)\).
The complete breaking pattern, including the internal symmetries is then
\begin{equation}
  SO(1,3) \times O(4) \to SO(3) \times D \times SO(2)^2 \to SO(2) \times D',
\end{equation}
with \(D' = D - \mu_1 SO(2)_1 - \mu_2 SO(2)_2\) (see also~\cite{Monin:2016jmo}).
The conformal group is broken to the sphere isometries \(SO(3)\) times the time translations \(D\), and the global \(O(4)\) is broken to \(SO(2)^2\) by fixing the charges.
This can be seen as an explicit breaking and does not give low-energy Goldstone \acp{dof}.
The second breaking is spontaneous, and the two broken generators \(T^a(\theta, \phi )\) of \(SO(3)\) produce only a single low-energy excitation.
This is because the equation~\cite{Low:2001bw}
\begin{equation}
  c_a(\phi) T^a(\theta, \phi) \varpi(\theta) = 0
\end{equation}
admits one non-trivial solution for the functions \(c_a(\phi)\).
All in all, we have three Goldstone \ac{dof}.
A study of the perturbation theory around \(\varpi(\theta)\) shows one field with linear dispersion relation \(\omega = c_s p\) (type-I Goldstone) and one with quadratic dispersion \(\omega \propto p^2\) (type-II Goldstone). Together, they account for all the \ac{dof}.
The speed of sound for the type-I Goldstone is fixed by three-dimensional scale invariance to be \(c_s = 1/\sqrt{2}\).
The low-energy excitations only propagate in the direction of the unbroken translation \(\phi\), so the corresponding Casimir energy is
\begin{equation}
  E_0 = %
  \frac{1}{2 \sqrt{2}} \zeta(-1/2|S^1) = \frac{1}{\sqrt{2}} \zeta(-1) = -\frac{1}{12 \sqrt{2}}.
\end{equation}

\subsection{Lattice Observables}
In order to establish that the discrete model described by the action in \cref{eq:fo4model} flows to the Wilson--Fisher $O(4)$ \ac{cft}, we have computed two observables: the current and the order parameter susceptibilities.

\begin{figure}
\includegraphics [width=0.8\linewidth]{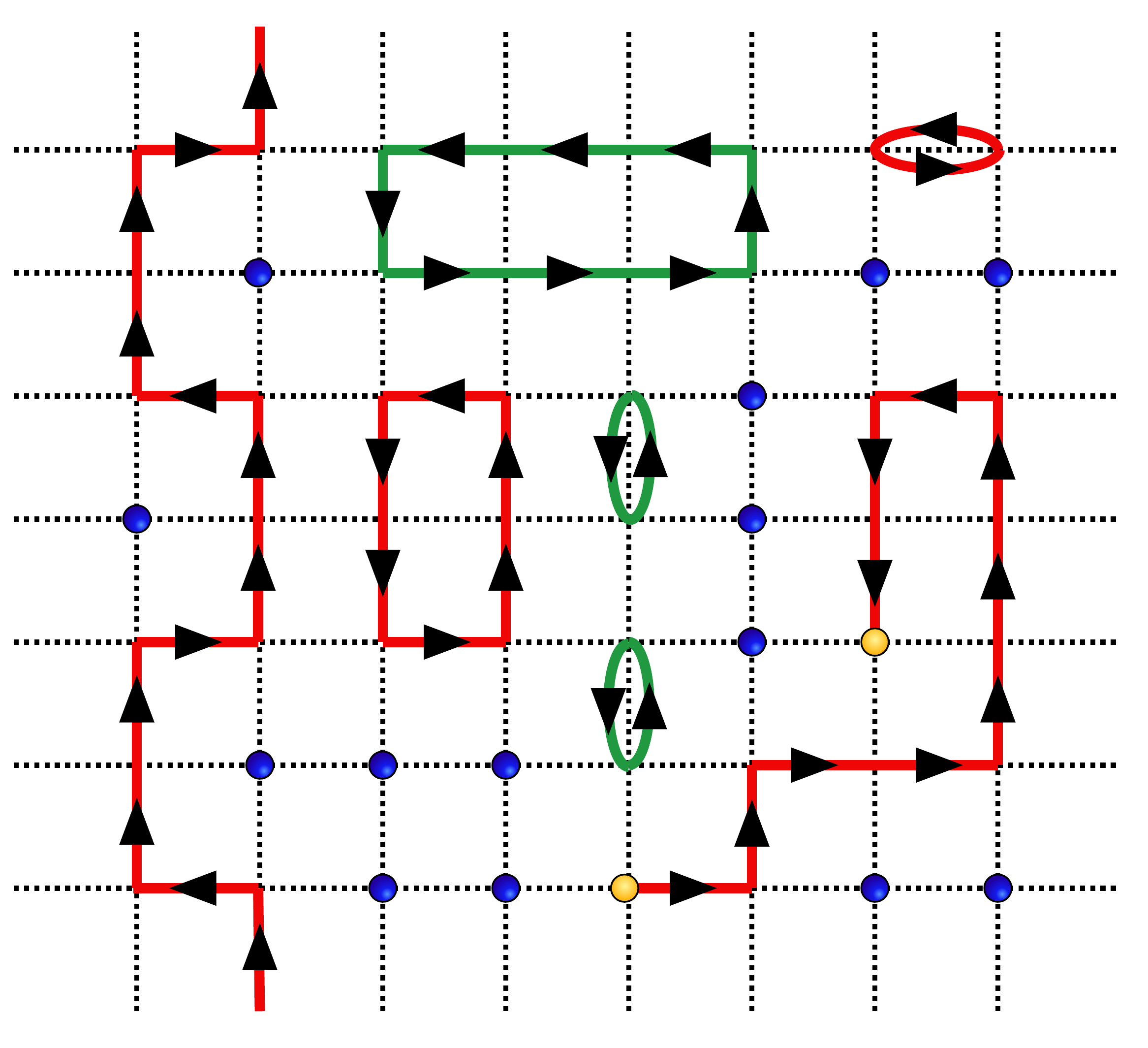}
\caption{Illustration of the $O(4)$ worldline configuration in two dimensions with a creation and annihilation event which contributes to Eq. \ref{eq:opsus} \label{fig:WLwS}}. 
\end{figure}

The current susceptibility is defined as
\begin{equation}
\rho_s = \frac{1}{L^3} \sum_{x,y} \ev{J_\mu(x) J_\mu(y)},
\end{equation}
where $J_\mu(x)$ is the $O(4)$ conserved current at the site $x$ in any fixed direction $\mu$.
In a worldline configuration, such as the one illustrated in \cref{fig:wlconf}, one can choose a surface perpendicular to $\mu$ and compute the current flowing out of that surface by counting the directed arrows of a given color with appropriate sign.
Red loops or green loops will give the same result due to the $O(4)$ symmetry.
The  order parameter susceptibility is defined as
\begin{equation}
  \chi  = \frac{1}{L^3} \sum_{x,y} \ev{ a_x a^\dagger_y}
  \label{eq:opsus}
\end{equation}
where $a^\dagger_y$ creates an $O(4)$ particle and $a_x$ destroys it. In terms of the worldline configurations, such creation and annihilation events are introduced as sources and sinks (see Fig \ref{fig:WLwS}) and can be sampled during the update of the worm (or directed loop algorithm) as described below. Hence, we can compute $\chi$ during such an update.

\subsection{Worm Algorithms}
Worm (or directed loop) algorithms are by now well established and easy to construct. Depending on the way the detailed balance is implemented, they can be constructed in different ways. Some algorithms can be found in the earlier work mentioned in the main paper. In this work we have used two different algorithms to make sure that each one is free of errors. When we compute the physical observables $\rho_s$ and $\chi$ in order to establish the $O(4)$ Wilson--Fisher fixed point, we used an algorithm which we refer to as \ALGOONE. This algorithm works in the absence of external charges (sources and sinks). We used a different algorithm, which we call \ALGOTWO, when updating configurations in the presence of sources and sinks. Below we give some details of \ALGOTWO. In fact, in the absence of the sources and sinks, \ALGOTWO can also be used to compute $\chi$ and $\rho_s$, and we have verified that they give the same numerical results as \ALGOONE within errors.
 
In the absence of the sources and sinks, \ALGOTWO begins in a configuration
that contributes to the partition function
but samples ``worm sectors'' with one additional source (tail) at $y$ and sink (head) at $x$ of a given color as illustrated in \cref{fig:WLwS}.  Each configuration in the worm sector is given a unique weight $W(x,y)$ depending on the location of the source and the sink. At the end of each worm update the source and the sink disappear and the configuration returns to the partition function sector.
The configurations generated during the worm update with a source and a sink help measure the order parameter susceptibility \cref{eq:opsus}. A brief description of the worm update is given below:
\begin{description}
\item[Begin] Choose a random initial site $y$ and a random color $c$ (red or green).
  \begin{enumerate}
  \item  If $y$ is a vacuum site, then propose to create the configuration in the worm sector with color $c$ with a probability $W(y,y)/U$. If the proposal is accepted based on a Metropolis accept/reject decision, create a configuration in the worm sector with both the head and the tail located on the site $y$.
  \item If $y$ contains a loop of color $c$, and this loop enters the site $y$ from $x$, break the loop between $y$ and $x$ and create a worm configuration with the tail at $y$ and the head at $x$ with a probability $W(x,y)$.
  \item If $y$ contains a loop of different color than $c$ or the proposal is rejected, then the worm update ends and no sources and sinks are created.
  \end{enumerate}
\item[Move] In the worm sector propose to move the head of the worm located at $x$, in any of the $d = 2 D + 1$ directions ($d = 0, \pm 1, \pm 2, \cdots, \pm D$) equally. Here $D=3$ is the dimension of the lattice. If we label the next site as $x_f = x + \hat{i}$ where $i$ is obtained from the direction $d$, the following possibilities can occur, assuming $x_f$ is not $y$: 
  \begin{enumerate}
  \item If there is a vacuum site at $x_f$, then propose to remove it with weight $W(x_f,y)/(U W(x,y))$. If accepted, move the worm head to the site $x_f$, otherwise keep the worm head at $x$.
  \item If there is a loop of a different color at site $x_f$, then keep the worm head at $x$.
  \item If there is a loop of the same color at the site $x_f$, then propose to merge the worm loop with the existing loop and move the worm head to the site $x_1$, from where the existing loop of the same color enters site $x_f$. This is a two site move of the worm head and is accepted with a probability of $W(x_1,y)/W(x,y)$. If the proposal is accepted move the head to $x_1$ otherwise keep the worm head at $x$.
    \item If the direction $d=0$ is chosen then propose to move the worm backwards to site $x_f$ by removing the bond at $x$ and creating a monomer on the site $x$. The proposal is accepted with probability $W(x_f,y) U/W(x,y)$.
\end{enumerate}
\item[End] During the move update if $x_f$ turns out to be the location of the tail $y$ or if the head and the tail are on the same site (\emph{i.e.} $x=y$) and the chosen direction is $d=0$, we propose to exit the worm sector. In the former case we propose to move the head to $y$ and close the loop and with probability $1/W(x,y)$. In the latter case we propose to create a vacuum site at $x$ with a probability $U/W(x,x)$. In both cases if the proposal is accepted the worm update ends, otherwise the worm head remains at its current location of $x$.
\end{description}
It is easy to verify that the above updates satisfy detailed balance at each step.
While in the worm sector, the worm loops contribute to $\chi$ when the weights $W(x,y)$ are appropriately divided out. On the other hand configurations in the partition function sector contribute to the observable $\rho_s L$.
Additional updates, such as changing the color of loops, as well as their orientation were used to achieve faster decorrelation.

\begin{figure}[t]
\includegraphics [width=0.8\linewidth]{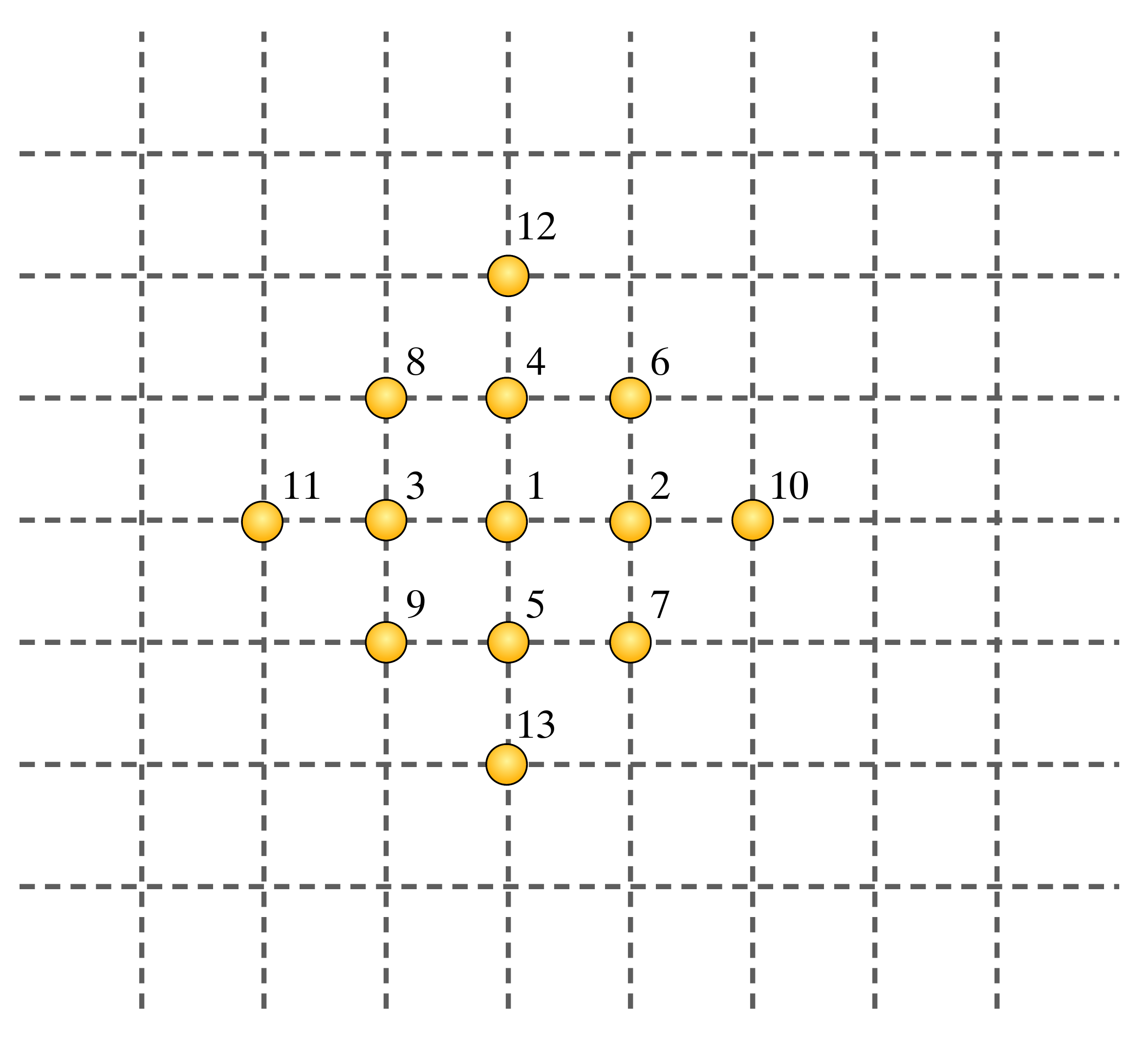}
\caption{One of many ways to create a source in the large charge sector. The circles represent the possible location of the sources and the numbers gives the order in which new sources are created to study the large $(j_{\Left}= j_{\Right}=j)$ sectors.}
\label{fig:srcsink}.
\end{figure}

The above algorithm can also be used to update worldline configurations in a fixed-charge sector as long as we make sure that the worm updates do not change the location of the sources and sinks. This is easily accomplished if we make sure that the initial site does not contain sources or sinks and if a worm move tries to change their locations it is not accepted. In this work we focus on sectors with charges ($q_{\Left}= q_{\Right}=j$).
Since in our lattice model we cannot create all the charges on a single site, we spread them around on a two-dimensional plane around a \emph{central site} on the $t=0$ and $t=L/2$ slices. One example of how we spread out the charges in the plane on both time slices is shown in Fig.~\ref{fig:srcsink}. The central site is labeled as $1$. For example, a charge with $j=3/2$ could be created with sources and sinks of red loops at sites labeled as $2,3,4$. Similarly, the charge with $j=3$ could be created with sources and sinks of red loops at sites $2,3,4,5,6,7$. We always leave out the central site for a reason that we explain below. We can define the sum over weights of such configurations with sources and sinks separated by $L/2$ in the sector with charge $j$ as the partition function $Z_j(L)$. We can then use our worm algorithm to sample the configurations that contribute to $Z_j(L)$. We can then also measure ratios of the form
\begin{align}
  R_j(L) = \frac{Z_{j}(L)}{Z_{j-1/2}(L)}.
\end{align}
For this purpose we first generate an equilibrium configuration that contributes to $Z_{j-1/2}(L)$. We then peform a special worm update called a measurement update that creates an additional $j=1/2$ charge (red loop) at the central site on the $t=0$ sheet. If this loop touches the central site on the $t=L/2$ we get a contribution to $R_j(L/2)$. In order to increase the efficiency of the measurement process we enhance the worm weights $W(x,y) \sim \abs{t_{xy}}^p$ so that the worms do grow large enough to contribute to the observable. This tuning is important and helpful when $j$ becomes large since there the ratio $R_j(L)$ is small.

We have checked that the algorithm can efficiently sample large $j$ sectors. We also find that the charges between the source and the sink flow either through the bulk, or over the boundary with equal probability. In addition, we have also checked that the choice of distribution of sources and sinks does not change the final results.

\subsection{Extraction of conformal dimensions $D(j,j)$}
\begin{figure}[t]
\includegraphics [width=\linewidth]{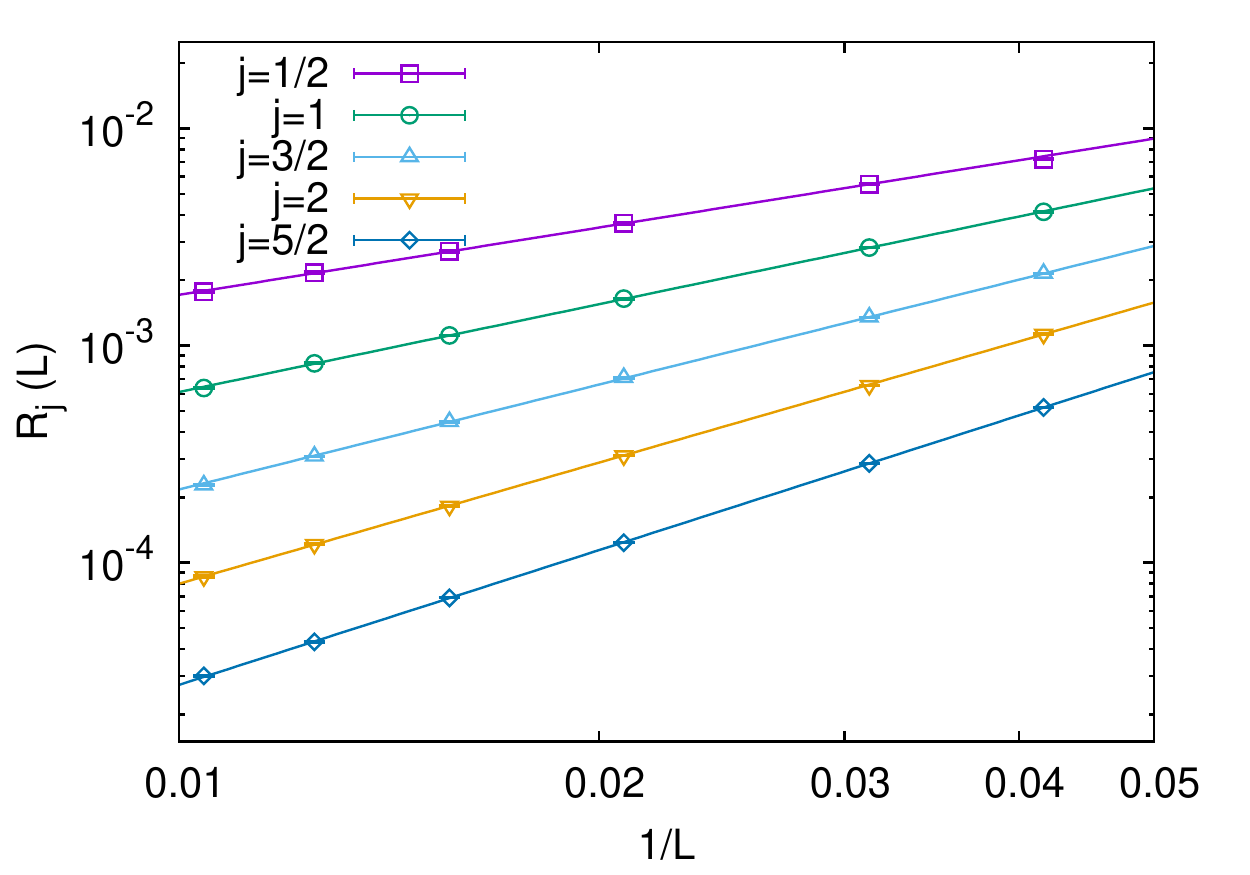}
\includegraphics [width=\linewidth]{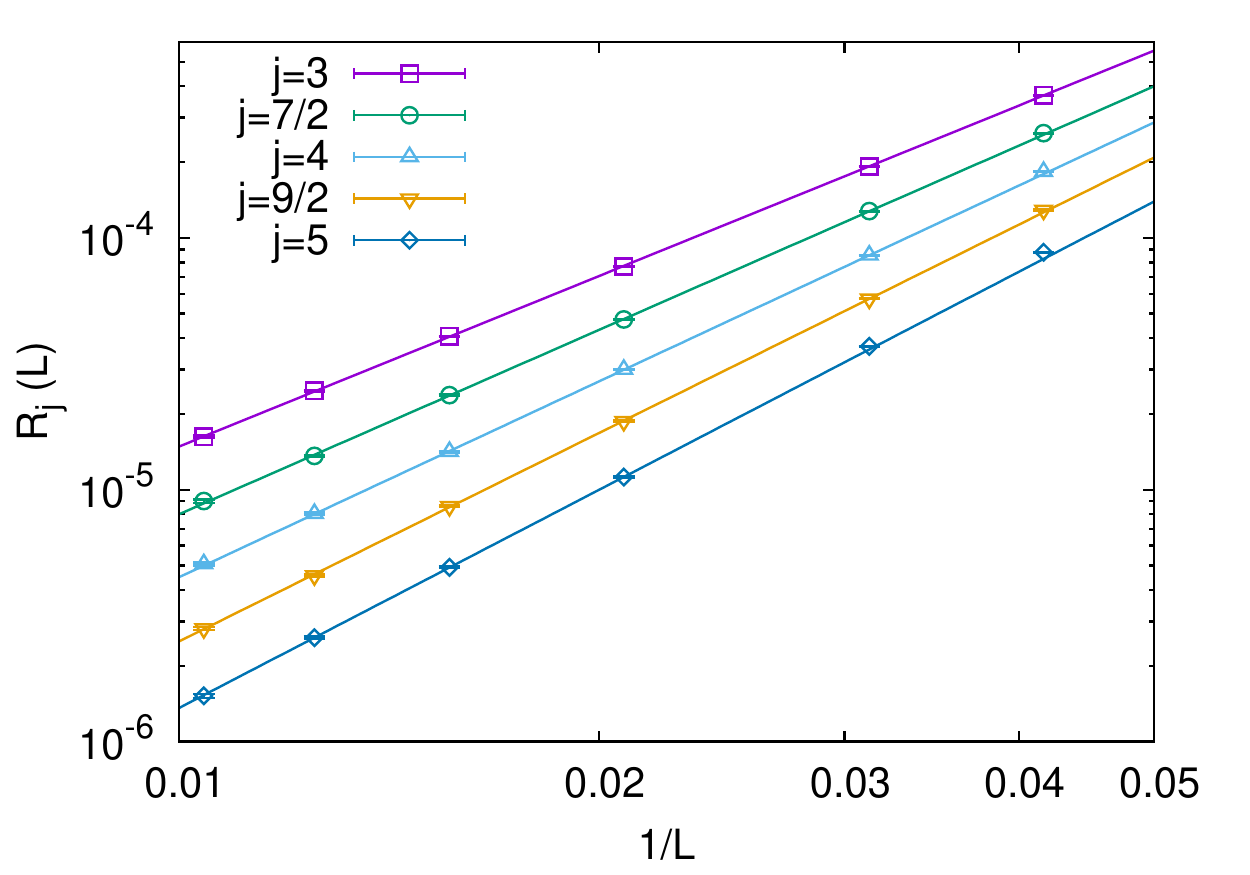}
\caption{The figures (top and bottom) show the quantity $R_j(L)$ for a range of lattice sizes $L/a = 24, \dots , 96$ and different representations $j_{\Left} = j_{\Right} = j$. The straight line fit on a log-log plot is indicative of the power law behavior, and the slope gives the difference of the conformal dimensions $2 \Delta(j)$. A very good $\chi^2/DOF \lesssim 1$ is obtained with a single power law for data points, indicating negligible scaling corrections, as well accurate extraction of the conformal dimension of the lowest operator in the $(j,j)$ sector. Due to the efficient worm algorithm, there is no visible signal-to-noise problem in these correlators. \label{fig:Rj}}
\end{figure}

In the main text, we have briefly stated how the idea initially developed in Ref.~\cite{PhysRevLett.120.061603} can be used to calculate the conformal dimensions $D(j,j)$ for the fermionic realization of the $O(4)$ model.
Here we provide further details. In the section above we explained how the measurement update  can be used to compute the ratio $R_j(L)$.
This observable gives us directly access to the conformal dimensions through the relation
\begin{equation}
  R_j(L) = \frac{Z_{j}(L)}{Z_{j-1/2}(L)} = \frac{C}{L^{2\Delta(j)}}\left( 1 + \frac{\gamma}{L^{\omega}} \right),
\end{equation}
where $Z_j(L)$ is the partition function in the charged sector with  $q_{\Left}= q_{\Right}=j$. At the critical point we expect $Z_{j}(L) \sim 1/L^{2 D(j,j)}$ which implies that $R_j(L)$ behaves as $C/L^{2\Delta(j)}$ for large lattices, where $\Delta(j)=D(j,j)-D(j-1/2,j-1/2)$. The term $\frac{\gamma}{L^{\omega}}$ are the scaling corrections, which are too small to be detected in our data for the lattices we have used. Hence, a single power law fit was adequate for the extraction of $\Delta(j,j)$. The results for $R_j(L)$ for various values of $j$ and $L$ are shown in Fig.~\ref{fig:Rj}.

The absolute conformal dimensions $D(j,j)$ are obtained using the relation
\begin{equation}
  \label{eq:DjjfromDeljj}
  \begin{aligned}
    D(j,j) ={}& [D(j,j) - D(j-1/2,j-1/2)] \\
    &+ [D(j-1/2,j-1/2) - D(j-1,j-1)] \\
    &+ \cdots + [D(1/2,1,2) - D(0,0)]\\
    ={}& \Delta(j,j) + \Delta(j-1/2,j-1/2) \\
    &+ \cdots + \Delta(1/2,1/2)
  \end{aligned}
\end{equation}
and using that $D(0,0)=0$. Moreover, since $\Delta(j)$ are extracted from fits to independent sets of simulations, there is no correlation between the different values of $\Delta(j)$. Hence, the errors in $\Delta(j,j)$ can be added in quadrature to obtain the error on $D(j,j)$. 

\bigskip

Using the extracted values of $\Delta(j)$ and $D(j,j)$, we perform a fit to the predicted conformal dimensions in the sector $j_{\Left} = j_{\Right} = j$ as given by formula~\ref{eq:dimensions-jj}. For this, we simultaneously fit the quantities $\Delta(j)$ and $D(j,j)$, keeping the value of $c_0 = -0.09372$ fixed. This yields the values of the coefficients to be $c_{3/2} = 1.068(4)$ and $c_{1/2} = 0.083(3)$. The systematic errors, as determined by changing the fit range exceeded the statistical error, and is the main source of the quoted errors. Although $c_0$ is known analytically, in order to determine how well our data is able to estimate this constraint, we performed a different fit where we allowed all the three coefficients $c_{3/2}$, $c_{1/2}$ and $c_0$ to vary.
The values of $c_{3/2}$ and $c_{1/2}$ obtained with this strategy are consistent with the previously quoted values, but the fit gives $c_0 = -0.072(10)$, which agrees with the theoretical estimate within two sigmas.
The $\chi^2/DOF$ in all the fits was around $0.01 - 0.20$, indicating the very good quality of the fits. The presence of the subleading terms in the expansion in Eq.~\ref{eq:dimensions-jj} is not observable in our calculations.
 
\end{document}